\documentclass[12pt]{article}
\usepackage{graphicx}
\begin{document}
\begin{center}
{\Large \bf Probing Very High Energy Prompt 
Muon and Neutrino fluxes and the Cosmic Ray Knee via Underground Muons}
\vspace{1cm}\\
Raj Gandhi \footnote{Email: raj@mri.ernet.in} and Sukanta Panda 
\footnote{Email: sukanta@delta.ft.uam.es, Present address: Dpt. de 
Fisica Teórica and Instituto de 
Fisica Teórica, Universidad Autonoma de Madrid, Madrid, Spain}\\
\hspace{10pt}\\
{\em Harish-Chandra Research Institute,}\\
{\em Chhatnag Road, Jhunsi, Allahabad 211019, India}.
\hspace{10pt}\\
\end{center}






\vspace{1cm}

\begin{abstract}
 We calculate event rates and demonstrate the observational feasibility of very 
high energy muons (1~TeV-1000~TeV) in a large mass underground detector 
operating  as a pair-meter. This energy range corresponds to surface muon 
energies of $\sim$(5~TeV - 5000~TeV) and primary cosmic ray energies of $\sim$ 
(50~TeV - 5 
$\times 10^4$ TeV). Such measurements would significantly assist in an improved 
understanding of the prompt contribution to $\nu_e, \nu_{\mu}$ and $\mu$ fluxes 
in present and future ultra-high energy neutrino detectors. In addition, they 
would shed light on the origin of and possible  compositional changes at and 
around  the observed 'knee' in the cosmic ray spectrum.
\end{abstract}




\section{The Cosmic Ray Spectrum and the Knee}

Cosmic ray studies, with the spectrum extending over ten decades in energy,
have proved to be fertile terrain for furthering our knowledge of both
astrophysics and particle physics ( reviews may be found in
 \cite {bere,gai,long,stan}). They have traditionally provided us with 
clues
 for the existence of new particles and the physics associated with them,
 which have later been confirmed by detailed accelerator experiments. In fact,
 prior to 1950 and the advent of modern accelerator technology, they
 provided the only means of studying  high-energy particle
 production and interactions. Additionally, as a result of 
our attempts to undestand the origin of cosmic rays, they have contributed
 to our knowledge of acceleration via shocks, and the propagation of charged
 particles in the galaxy and heliosphere.

The cosmic ray spectrum, characterised by a steeply falling power-law
behaviour over its entire range,  exhibits  two transition regions where
the slope changes noticeably:
\begin{itemize}
{\item A steepening of the  spectrum occurs around $E\approx 5 \times 
10^6$ GeV, ${\it
  i.e.}$ the  index $\gamma$ describing  the power-law
  behaviour of the differential flux, $dN/dE \sim E^\gamma,$ 
changes from $\gamma\approx -2.7$ to
  $\gamma\approx -3.1$; leading to the feature called the `knee'.}

{\item A flattening  of the  spectrum occurs around $E\approx 5 \times 
10^9$ GeV, {\it i.e} at the ``ankle''; with the  index $\gamma$ changing back to
  $\sim 2.4-2.7.$ Beyond the ankle, in the realm of ultra high energy cosmic
  rays, data \cite {nag,hir,aga} is sparse and conflicting, but highly
  intriguing. While we will not address the interesting puzzle in this regime
   here, a discussion of the various issues  may be found in \cite {pij,hal1}, 
and a recent assesment of the shape of the spectrum based on current 
knowledge can be found in \cite {stan2}.}
\end{itemize}

The physical reason for the existence of the knee is at present an 
unresolved problem of great significance to understanding the origin of
galactic cosmic rays. It is generally believed that the reasons underlying
this distinctive shift in the spectrum are astrophysical in nature, as
opposed to those stemming from a change in hadronic interactions at these 
energies which, at present, are not within the reach of existing accelerators.  
This conclusion is based on the observed correspondence between independant
measurements of the muon number spectrum, Cerenkov radiation and hadronic
constituents of air-showers \cite {mb,kas1,tun,kas2,agl2}. While the 
reasons for the shift in the spectrum remain un-understood, these data exhibit an
expected co-relation which supports the 
absence of  radically different physics interactions at these energies. 

While the case for the existence of new physics being at least 
partially responsible for a shift in the spectral index   is not 
wholly without motivation\footnote{This region in $E_{p}$, (i.e, the
energy of the primary particle) corresponds to several TeV in center of mass 
energies. Thus  there are many conjectures for physics beyond
the Standard Model which come into play, {\it e.g.}  SUSY, technicolour, large extra dimensions
etc. These could lead (via new particle production and decay) to energy being
channelled into muons, neutrinos or other secondary particles in a
manner that present cosmic ray experiments are insensitive to, causing the
 shift in the (measured) energy spectrum \cite {nik,pet1,kaz1,kaz2}.
}, we stress that it appears unlikely that this can be empirically corraborated
or refuted in the near future in CR measurements. This is because, 
as we shall discuss below, uncertainties (in the knee region) in 
the CR compositon and prompt muon and neutrino contributions would 
likely overshadow evidence of such new interactions. 
 In any case, this hypothesis will
 be thoroughly probed in the near future by the Large Hadron Collider (LHC) 
at CERN which will operate at a center of mass energy of $\sqrt{s} = 14$ TeV. 

\section{Uncertainties in the Muon and Neutrino Fluxes in the Knee region and beyond}

As stated earlier,  present data \cite {hor2} when culled 
and correlated, appear to favour one or more  astrophysical reasons for the 
existence of the knee. These include it being a rigidity-dependant 
effect (originally proposed in {\cite {pet}}) related to the (different)
maximum acceleration energies for different nuclei 
either in the cosmic ray source itself 
or during the propagation process. Data from surface air-showers and 
optical detectors indicate, without
being conclusive, that the average mass of the cosmic ray spectrum
nuclei differs before and after the steepening at the knee. In particular, there appears to be 
some evidence\cite{kas3,eas2} that the composition is heavier above the knee region. 
If this is true, then, as discussed in \cite{can}, 
significant suppression of the very high energy ($\geq 10^5$ GeV) muon and   
neutrino fluxes resulting from CR interactions in the 
atmosphere and in the interstellar medium can occur.  

A second major factor in determining the enhancement (or lack thereof) of 
muon and neutrino fluxes above several TeV are the uncertain magnitudes of the 
prompt ({\it i.e.} those resulting from heavy meson decay, notably, charm mesons and 
heavier composites) fluxes of both. At  low ({\it i.e.} $\sim$ GeV) energies, the
cosmic ray induced neutrino and muon fluxes receive their dominant contributions
from the decays of $\pi$ and $K$ mesons, whose interaction lengths significantly  
exceed their decay lengths\cite{vol, gai1,lip,tig}. 
( These fluxes are henceforth referred to as the conventional fluxes in what follows.)
This sitiuation changes  at $\sim$ TeV energies, and secondary interactions of
these particles become possible,  leading to the production of heavy short lived
hadrons.  While  upper bounds on
the flux of  muons and neutrinos have been provided by several experiments {\it e.g} LVD {\cite{lvd}}, 
AKENO {\cite{ake}} and AMANDA {\cite{aman}}, they
still allow for a very large possible range of prompt  flux magnitudes.  

Present phenomenological predictions for the diffuse fluxes of these prompt 
muons and neutrinos can differ  by about two orders
of magnitude {\cite{tig,zhv,rvs,prs,bnsz,ggv2, mrs}}. The sources of this
large uncertainty lie, to a significant extent, in the choice of charm 
production models. For instance, differing predictions arise from models
based on perturbative QCD (pQCD) with a $K$ factor {\cite{tig}}, 
next-to-leading order (NLO) pQCD {\cite{ggv2,prs}}, 
quark-gluon string models and recombination quark-parton
models{\cite{bnsz} etc. In general, QCD based models must 
contend with a large uncertainty associated
with the extrapolation of the gluon parton distribution function
$g(x)$ to small fractional momentum $x < 10^{-5}$. Theoretical models generally 
assume $$ xg(x) \sim x^{-\lambda},$$ where $\lambda$ is in the range 
$0-0.5$, and fluxes depend strongly on  the chosen value of $\lambda$. We 
note that depending on the model, the prompt muon and neutrino fluxes from charm
decay exceed the corresponding conventional fluxes (from $\pi$ and $K$ 
decays) somewhere between (surface) muon energies of few tens of TeV and 
few PeV \cite{ggv2}. Reliable measurements of muon fluxes in this range would 
thus, at the very least, help in establishing the reliability of a particular class of models.

 While we do not give a detailed  account of
the flux predictions from all the different models, we attempt to
give a representative idea in our calculations of the variation possible even 
within a given charm production model. 
While the conventional muon flux from $\pi$ and $K$ decays is well understood 
and fairly firm, the prompt flux
predictions are subject to  variations resulting from different parton
distribution functions and choices of the factorization and
renormalisation scales as mentioned above. We thus  
stress the need for better empirical determination of 
the muon (and associated neutrino fluxes) in this region, 
a topic which is elaborated upon in the next section.

\section{The significance of measurements of  the muon and neutrino fluxes in the knee region}
 
With very few exceptions, available data on 
muons above several TeV comprise of measurements of the 
number spectrum rather than  the energy. This is 
primarily due to the size and density requirements imposed 
on detectors by the significant penetration 
lengths acheived by high energy muons.

The desirability of improved and statistically significant 
muon energy measurements in the few TeV to few hundred TeV 
region stems from (at least) three reasons:

\begin{itemize}
{\item As mentioned above, the physical origin and composition of 
cosmic rays in this energy range is currently 
obscured by a paucity of data on VHE  muon and neutrino  fluxes. 
 Observations would thus certainly illuminate the current debate on the 
reason for the occurance of and compositional changes at the  knee.}

{\item Also, as discussed above, QCD related  theoretical uncertainties 
dominate the predictions of the prompt contribution to muon and 
neutrino fluxes. As emphasized in \cite{ggv2,ggv}, the measurement 
of down-going muon fluxes would provide a valuable handle in their 
reduction.}

{\item Both the conventional and prompt  muon fluxes at these  energies 
are closely related to the associated neutrino fluxes. For 
prompt contributions, this is is because the kinematics of 
charmed particle decay and the corresponding semi-leptonic 
branching ratios ensure that the $\nu_e$ and $\nu_{\mu}$ fluxes 
are identical upto a few percent to the 
muon fluxes in this energy range, regardless of the choice of the charm
production model or of $\lambda.$ The conventional 
neutrino flux, on the other hand, is about 10 \% of 
the conventional muon flux. At the energies of 
interest here, neutrinos resulting from cosmic ray 
interactions in the atmosphere and in the inter-stellar 
medium  constitute the most important background to 
searches for diffuse fluxes of  ultra high energy (UHE) 
neutrinos \cite{ bere2,sig2,rg1,rg2,rg3,gai2,hal2} from cosmological
sources (${\it e.g.}$ active galactic nuclei, gamma-ray bursts etc.) in 
neutrino telescopes like AMANDA \cite{sil}, ICECUBE \cite{tdy}
and NEMO \cite{cir}. Thus, they are an important 
obstacle to the much-anticipated detection of such 
energetic point sources in these detectors. Empirical 
data on downgoing muons from cosmic rays would 
prove invaluable in understanding this background 
to the UHE neutrino signal, since it can be co-related to the neutrino flux.}   
\end{itemize}
In the context of the points above, it is relevant to 
stress the importance of being able to dis-entangle 
the prompt (due to the decay of produced heavy mesons) and 
conventional (from $\pi$ and $K$ decays) and diffuse 
UHE (from extra-galactic sources) contributions  to 
the neutrino fluxes. Methods to enable this 
have been studied, and are based  on the differences in 
zenith angle, depth and spectral  dependance 
between these fluxes \cite{sine1,sine2} and 
on the isolating capability of showering  $\nu_e$ 
charged current events via a break in the spectrum \cite{bea}.

Having emphasized the importance of muon energy measurements in 
the several TeV to several hundred TeV range,  we proceed in the next section
to study the potential of the {\it pair meter } method 
{\cite{alek,kp,kp1}} as applied to such 
measurements made in a {\it large iron calorimeter} (50 kT) 
\footnote{Such a detector is currently being planned 
for location in India \cite{ino}}.
Since individual muon energies will
become measurable using this technique, it will be
possible to augment the sparse existing data on cosmic ray muons in the 
important  range where they have {\it surface} energies  
of $ \approx 5-5000$ TeV. Furthermore, these observations can be 
 combined with balloon-based experiments ({\it e.g} TRACER \cite{mul}) and 
upcoming hybrid air-shower experiments ({\it e.g} 
KASCADE-Grande  \cite{nav} and 
LOPES \cite{fal}) to enhance our understanding of the issues 
discussed above.  We mention here that this range in {\it muon surface energy}
roughly corresponds to a range of $50-5\times 10^4$ TeV in {\it primary cosmic 
ray energy}, which is crucial to an enhanced understanding
of the origin of the knee.  
 
\begin{figure}[t]
\hbox{\hspace{0cm}
\hbox{\includegraphics[scale=.8]{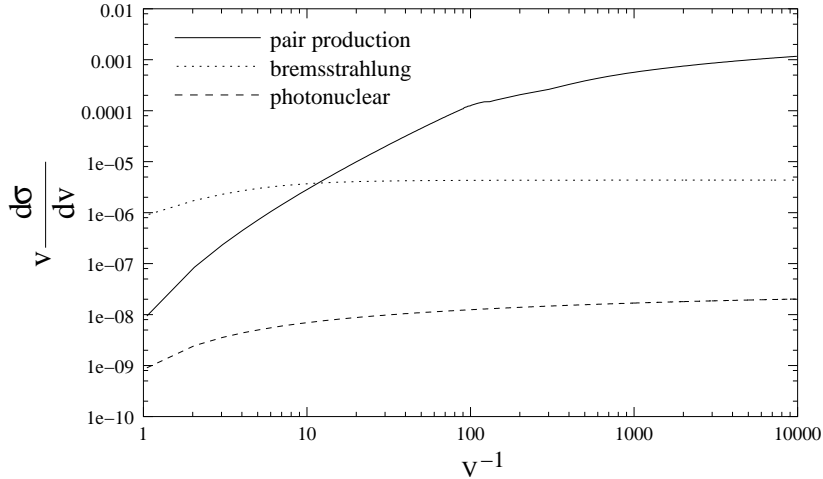}}}
\caption{Differential cross section $v d\sigma/dv$
vs. $v^{-1}$(inverse of the relative energy transfer)for pair
production(solid)\cite{ker}, bremstrahlung(dotted)\cite{ps} and
photonuclear(dashed)\cite{bb} processes.}
\label{dr}
\end{figure}

In the remainder of the paper, we first provide a discussion of
the pair-meter technique  and the pair production cross section 
which results in the observed cascades. This is followed by a brief 
description of a typical
large-mass iron calorimeter. Subsequent to this we summarize the interactions
and losses of muons in matter enroute to an underground detector, and their
incorporation into our calculation.   We
then calculate anticipated event rates for a 50 kT detector and
demonstrate that even after accounting for energy losses in the surrounding rock, 
event rates can be appreciably large for the $1-1000$ TeV range, corresponding 
to {\it surface} muon energies in the range of several  TeV to several PeV.  
%

   
\section{The Pair Meter method and the associated Pair Production Cross Section} 

Due to the penetrating power of muons, their energy measurements 
require  techniques which differ from those
employed for photons, hadrons and electrons. Furthermore, muon energy
measurement methods which work well in the GeV range (magnetic spectrometry
or measuring Cerenkov radiation) are rendered impractical in the TeV range
primarily due to  requirements of size imposed by the combination
of high energies and a steeply falling spectrum.  

The pair meter technique{\cite{alek,kp,kp1} skirts some of  the 
disadvantages of traditional muon detectors by  relying on a somewhat 
indirect method, {\it i.e.} the measurements of the energy
and frequency of electron-positron pair cascades produced 
by the passage of a high energy muon in dense matter. A reliable 
reconstruction of the muon energy in this method is based on
the following:

\begin{itemize}
\item {The cross section for $e^+e^-$ pair production by a muon with 
energy $E_{\mu}$ with energy transfer
above a threshold $E_0$ grows   as
$ln^2(2m_eE_{\mu}/m_{\mu}E_0)$, where $m_{\mu}$ and $m_e$ are the 
muon and electron masses respectively.} 

\item{ Defining $v=E_0/E_{\mu}$,  above  $v^{-1}=10$, this cross
section  dominates those for other muon energy loss processes which generate
observable cascades in its passage through dense matter, {\it e.g.} $\mu-N$
inelastic scattering and bremstrahlung emission. This is demonstrated
in Figure \ref{dr}, where  we compare the 
differential cross sections for these various interactions as a function of 
$v^{-1}$.}

\item{The energy lost to each cascade resulting from  
$e^+e^-$ pair production
is a very small fraction (about $10^{-2}$) of the muon energy for the
range of $v^{-1}$ which we focus on here.}

\item {The dependance of the pair production cross section on 
$E_\mu/E_0$ then allows one to infer the muon energy by counting the number of 
interaction cascades N in the detector with energies above a threshold $E_0$.} 
\end{itemize}
We now  make the above statements more precise.  In the approximations 
$$v=\frac{E_0}{E_\mu} \ll \frac{2 m_e}{m_{\mu}}$$\\ 
and $$E_0 \gg 2 m_e 189 \sqrt{e} Z^{-1/3} \simeq 0.3 Z^{-1/3} GeV,$$ 
(Z=atomic number= 26, for iron) both of
which are valid for the choice of $E_0, E_{\mu}$ for which we present results 
below, the expression for the 
differential pair production cross section is given by{\cite{ker}}  

\begin{equation}
v \frac{d\sigma}{dv}  \simeq \frac{14 \alpha}{9 \pi t_0} 
ln\left( \frac{\kappa m_e E_\mu}{ \epsilon m_{\mu}} \right) \ ,
\end{equation}
where $\alpha = 1/137$ and 
$\kappa \simeq 1.8$. $t_0$ is the radiation lenth (r.l) which is given by 
\begin{equation}
t_0 = \left(\frac{4 Z (Z + 1)}{A_W}~ \alpha r_0^2 N_A~ ln(189 Z^{-1/3})  
\right)^{-1} \ .
\end{equation} 
Here $A_W$ is the gram atomic weight (for iron, this is 56 grams), $r_0$ is the classical electron radius and 
$N_A$  the Avogadro number. For iron, this gives $t_0 = 13.75$ gm/cm$^2$.

The average number of interaction cascades $M$ above a threshold 
$E_0$  for $v\leq 10^{-3}$ is given by 

\begin{equation}
M(E_0,E_\mu) = Tt_0\sigma(E_0,E_\mu) \ ,
\label{ni}
\end{equation}
where T is the thickness of the target in units of $t_0$ and
$\sigma(E_0,E_\mu)$
is the integrated cross section (in units of cm$^2$/gm), 
\begin{equation}
\sigma (E_0, E_\mu) \simeq \frac{7 \alpha}{9 \pi t_0}
\left(ln^2\left( \frac{\kappa m_e E_{\mu}}{ E_0 m_{\mu}} \right) + 
C\right) \ ,
\label{cros}
\end{equation}
where $C \simeq 1.4.$

 At this point it is important to mention that the capability and effectiveness of the 
pair meter method for high energy muons has been tested and demonstrated by the NuTEV/CCFR collaboration,
as described in \cite{apc}.
The calculations which follow are performed for a
50 kT  iron calorimeter. Our prototype is based on the suggested
design for INO; see {\cite{ino} for details. The dimensions of a 50 kT
detector of this type would correspond to (approx) 15 m $\times$ 15 m
$\times$ 45 m. A muon traversing a 20 m path in this detector
corresponds to a path-length of $\sim 1145$ r.l. In what follows, we
assume a (conservative) ``average'' path-length of 1000 r.l for the
typical muon and calulate the number of observable cascades produced
by it, for different cascade thresholds and  muon energies. Figure \ref{noi}
shows the average number of cascades above a threshold energy $E_0$
produced by a muon entering the detector with energy $E_{\mu}$ and
$T=1000$ r.l.; for three different choices of $E_0$, {\it i.e.} 1 GeV,
10 GeV and 100 GeV.
\begin{figure}[htp]
\hbox{\hspace{0cm}
\hbox{\includegraphics[scale=.8]{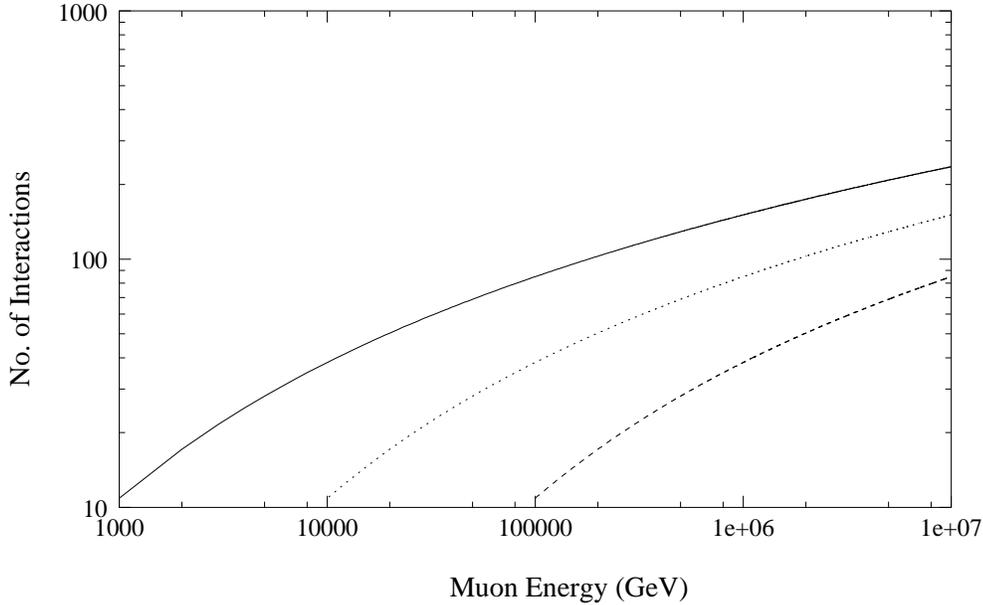}}}
\caption{Average number of cascades above a threshold $E_0$  vs. muon energy
for $E_0 = 1$ GeV (solid line),
10 GeV (dotted) and 100 GeV (dashed), with T  fixed to 1000 r.l.}
\label{noi}
\end{figure}
  Quantitatively, we note that
this leads to a $E_\mu=100$ TeV muon  generating approximately 40 cascades, each
of energy greater than $E_0=10$ GeV and 10 cascades with energy in excess of
100 GeV. By counting the cascades for several choices of 
thresholds for a traversing muon, one obtains a reliable estimate of its
energy. 

It is also relevant to remark here  that the relative energy
measurement error, $\delta E_\mu/E_\mu$ in the pair meter  is
given by
\begin{equation}
\delta E_\mu/E_\mu = \sqrt{\frac{9\pi}{28\alpha T}}\simeq\sqrt{\frac{137}{T}} \ .
\end{equation}
For $v=(10^{-3} - 10^{-2})$, which is the range we focus on here, this allows a liberal tolerance for error in the
measurements of individual cacade energies. We note also that the
errors do not worsen with increasing muon energy, which is an
important advantage of the pair-meter technique.

\section{The Surface Muon Energy determination for Underground Events}

It is  important to co-relate the measured muon energies in an underground 
detector to their surface energies, which we take to be those that would be 
observed were our detector placed on the surface of the earth. This
requires a calulation of the energy loss as the muon traverses the rock
between the earth's surface and the detector.

 These losses originate from 
ionization, bremsstrahlung, pair production and photonuclear interactions.
They can be effectively parametrized {\cite{ker, ps, bb}} for
 $E_{\mu} \geq $ 1 TeV, since  the average loss  increases
predominantly linearly with energy,

\begin{equation}
\left<\frac{dE}{dX}\right> = - \alpha - \beta E \ ,
\end{equation}
where $\alpha$ parametrizes the  contribution from ionization of muons and $\beta$
encapsulates the contribution from bremsstrahlung, pair production and photonuclear
processes.  Note that  $\alpha$ and $\beta$ carry a very weak(intrinsic) energy 
dependance.  It
is thus  appropriate to assume that the 
average muon energy at depth $X$ is
\begin{equation}
\left<E_\mu(X)\right> = \left(E_{\mu}^s + \frac{\alpha}{\beta} \right) 
e^{- \beta X} - \frac{\alpha}{\beta} \ ,
\label{loss}
\end{equation}
where $E_\mu^s$ is the initial surface  muon energy. One may use this
 to write down the  minimum surface energy required of a muon 
to reach a depth $X$ as, 
\begin{equation}
E^s_{min} = \frac{\alpha}{\beta} \left( e^{\beta X} - 1 \right) \ .
\end{equation}

 From Eq. \ref{loss}, we get the relation between initial energy $E_{\mu}^s$
and degraded energy of muon $E_{\mu}$ after travelling a distance $X$ 
as,
\begin{equation}
E_{\mu}^s = \left(E_{\mu}+\frac{\alpha}{\beta}\right) e^{\beta X} - 
\frac{\alpha}{\beta} \ .
\end{equation}
The differential muon flux at a depth $X$ is given by,
\begin{equation}
\frac{dN}{dE_{\mu}} = \frac{dN}{dE_{\mu}^s} e^{\beta X} \ .
\end{equation}  
where $\frac{dN}{dE_{\mu}^s}$ is the initial muon flux 
with surface muon energy $E_{\mu}^s.$
\begin{figure}[htp]
\hbox{\hspace{0cm}
\hbox{\includegraphics[scale=.9]{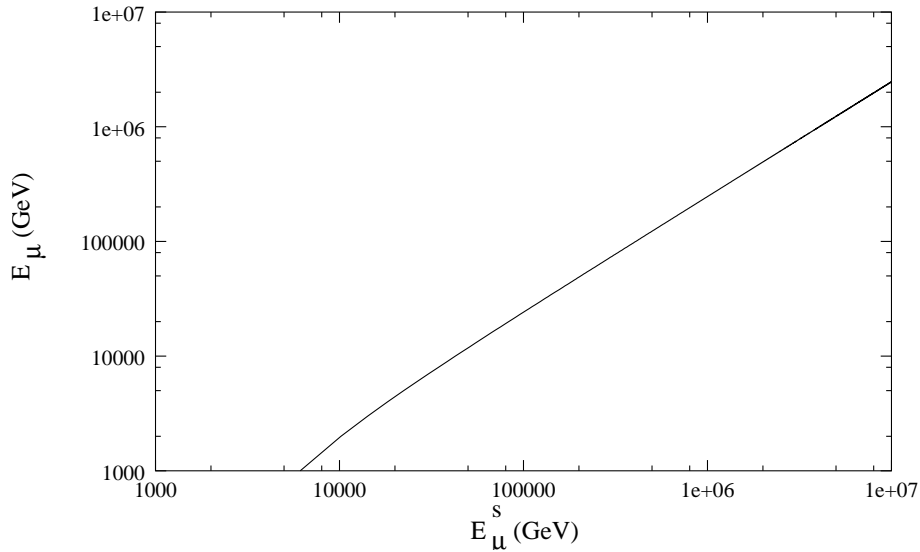}}}
\caption{Degraded muon energy, $E_{\mu}$ vs. surface muon energy, $E_{\mu}^s$ 
after passing a depth
of $3.5 \times 10^5 g/cm^2,$ corresponding to the proposed INO location.}
\label{fluxf}
\end{figure}

 The $\alpha$ and $\beta$ in the 
analytical expression can be obtained for standard rock. The depth
relevant to the INO detector's proposed location is 
$3.5 \times 10^5$ gm/cm$^2.$ The values are, 
\begin{equation}
\beta = 4 \times 10^{-6} cm^2/g,~ \frac{\alpha}{\beta} = 675 GeV.
\end{equation}
Figure \ref{fluxf} shows the degraded muon energies ({\it i.e.} those measured
for muons entering the detector after traversing the rock)  vs 
their corresponding surface energies after losses are accounted for in the manner 
described above. We note that typically, $E_{\mu}^s \simeq (2-5)\times E_{\mu}.$

\begin{table}[h]
\begin{tabular}{|l|l|l|l|l|l|l|l|r|}
\hline
Flux model & PDF & Scales  & a & b & c & d\\
\hline
PRS1 & CTEQ3 & $\tilde{M}=\tilde{\mu}=m_c$ & 5.37 & 0.0191 & 0.156 & 0.0153 \\
PRS2 & CTEQ3 & $\tilde{M}=2\tilde{\mu}=2m_c$ & 5.79 & 0.345 & 0.105 & 0.0127 \\
PRS3 & D & $\tilde{M}=2\tilde{\mu}=2m_c$ & 5.91 & 0.290 & 0.143 & 0.0147 \\
\hline
\end{tabular}
\caption{PRS  parameters  for the prompt muon and antimuon fluxes. $m_c$ is the
mass of the charm quark.}
\label{t}
\end{table}
\begin{table}[htp]
\begin{tabular}{|l|l|l|l|l|l|l|l|r|}
\hline
\multicolumn{1}{|c|}{} &
\multicolumn{6}{c|}{Number of muons per solid angle entering the detector
in 5 years}
\\
\hline
$E_{\mu}(TeV)$ & conv+TIG & conv & TIG & PRS1 & PRS2 & PRS3 \\ \hline
  1 & $1.035\times10^7$ & $1.03 \times 10^7$ & 37461 & 55482 & 95489 &
136871\\
  10 & 52486 & 51282 & 1204 & 2952 & 5341 & 10443 \\
  50 & 770 & 696  & 74 & 236 & 431 & 1104 \\
  100 & 127 & 106 & 21  & 73 & 134 & 387 \\
  200 & 22  & 16  & 6   & 22 & 40 & 129 \\
 300 & 8   & 5  & 3 & 11  & 19 & 66 \\
  400 & 4   & 2  & 2   & 6  & 11  & 41 \\
  500 & 2   & 1 & 1  &  4  & 7  & 28 \\
  600 & 1.5   & 1 & .5  &  3  & 5  & 20 \\
  700 & 1  & .5& .5  &  4  & 7.5  & 31  \\
  800 & .8  &.35 & .5 &  1.5  & 3  & 12  \\
  900 & .65 & .25 & .37  & 1.25 & 2.5  & 10 \\
  1000 &.5 &  .2  & .3  & 1  & 2 & 4 \\
  10000& .0025 & .0003& .0022 & .007 & .013 & .08 \\
\hline
\end{tabular}
\caption{Number of muons per solid angle entering the detector over  5
years
for various energies of the entering muon, $E_\mu$. }
\label{tab3}
\end{table}
\begin{table}[htp]
\begin{tabular}{|l|l|l|l|l|l|l|l|l|l|l|l|r|}
\hline
\multicolumn{2}{|c|}{} &
\multicolumn{9}{c|}{Number of cascades per muon for different thresholds $E_0$ 
in GeV} \\
\hline
$E_{\mu}$ & $E_{\mu}^s$ & 5 & 10 & 20  & 50 & 100 & 300 & 
500 & 1000 & 5000\\ 
\hline
  1 & 6.1 & 3.08 & 2.56 & 3.78 &  & & & & &\\
  10 & 40.26 &17.28 & 10.99 & 6.43& 3.08 & 2.56 & & & &   \\
  20 & 83.16 & 25.3 & 17.28  & 10.99 & 5.34 & 3.08 & & & &\\
  50 & 205  & 38.58 & 28.26  & 19.67 & 10.99 & 6.43 & 2.78 & 2.56& &\\
  100 & 407.58 & 50.63 & 38.58 & 28.26  & 17.28 & 10.99 & 4.58& 3.08& 2.56&\\
  200 & 813 & 64.43 & 50.63 & 38.58  & 25.30  & 17.28 & 8.11& 5.34& 3.08&\\
  300 & 1218 & 73.3 & 58.49 & 45.42  & 30.8  & 21.76  & 10.99 & 7.46 & 4.19 &\\
  400 & 1624 & 79.96   & 64.43 & 50.63  & 35.06  & 25.3 & 13.39 & 9.33 & 5.34 & 
\\
  500 & 2029 & 85.33  & 69.24 & 54.89  & 38.58  & 28.26  & 15.45 & 10.99 & 
6.43&2.56 \\
  600 & 2435 &89.85  & 73.3 & 58.49  & 41.58  & 30.8  & 17.28& 12.47& 
7.46 &2.58\\
  700 & 2841 & 93.76 & 76.83 & 61.64& 44.21  & 33.05  &18.91 &13.82 &8.43 
&2.6\\
  800 & 3246 &97.23  &  79.96 & 64.43  & 46.56  & 35.06  & 20.4& 15.06& 
9.33 &2.72\\
  900 & 3652 &100.33  & 82.77 & 66.95  & 48.69 & 36.9  & 21.76& 16.21& 10.18& 
2.89 
\\
  1000 &4057 &103.16  & 85.33 & 69.24  & 50.63 & 38.58  & 23.02 & 17.28 & 10.99 
& 3.08\\
  10000 & 40554 & 174.84  & 151.24 & 129.38 & 103.16 & 85.33 & 60.63 & 50.63 & 
38.58 & 17.28 \\
\hline
\end{tabular}
\caption{Number of cacades above thresholds $E_0=5, 10, 20, 50, 
100, 300, 500, 1000, 5000$ GeV per muon. Here
$E_\mu$ is the energy of the muon in TeV entering the detector, and $E_\mu^s$
is its corresponding energy in TeV at the surface of the earth, assuming it
traversed a depth of rock corresponding to $3.5 \times 10^5 gm/cm^2.$}
\label{tab1}
\end{table}

\section{Muon Fluxes}
 Extensive predictions and studies \cite{vol,
 gai1,lip,tig,zhv,rvs,prs,bnsz,ggv2, mrs} for prompt  cosmic ray muon fluxes at very
 high energies exist in the literature, as mentioned earlier. For 
our representative calculations of muon event
 rates, we have used the relatively conservative predictions for charm
 induced fluxes given in
 {\cite{tig,prs}}. The large variation in muon rates possible due to
 flux uncertainties  even
 when  these fluxes are used is amply reflected in our results, most noticeably
in Table 2. One would expect much larger variations if the full range of
 prompt flux models available is used to calculate event rates.

In {\cite{tig}} (henceforth referred to as the TIG flux), the conventional 
and prompt fluxes have been parametrized as
\begin{equation}
\frac{dN}{dE} = \frac{N_0 E^{-\gamma - 1}}{1 + A E} 
\end{equation}
for $E < E_{a}.$ and as 
\begin{equation}
\frac{dN}{dE} = \frac{N_0^{'} E^{-\gamma^{'} - 1}}{1 + A E}
\end{equation}
for $E > E_{a}.$
For the conventional muon flux $N_0 = 0.2,$ $N_0^{'} = 0.2,$ $\gamma=1.74,
$ $\gamma^{'} = 2.1,$ $E_a = 5.3 \times 10^5,$ $A = 0.007.$

For the prompt muon flux $N_0 = 1.4 \times 10^{-5},$ $N_0^{'} = 4.3 \times 
10^{-4},$ $\gamma=1.77,
$ $\gamma^{'} = 2.01,$ $E_a = 9.2 \times 10^5,$ $A = 2.8 \times 10^{-8}.$

The second set of representative  prompt muon fluxes we use are
calculated in \cite{prs} (henceforth referred to as the PRS1,PRS2 and
PRS3 fluxes).
The   differences in the three  fluxes originate in  
 different choices of parton distribution functions(PDF) and 
factorisation ($\tilde M$) and renormalisation scales($\tilde{\mu}$) of the 
theory.
These fluxes can be convieniently parametrized {\cite{prs}} as follows  
\begin{equation}
\frac{dN}{dE} = 10^{-a + b x + c x^2 - dx^3}, 
\end{equation}
where $x= Log_{10}$(E/GeV), with   $a,b,c$ and $d$ as  in 
Table.\ref{t}.  

In Figure \ref{flux1} we show the conventional (TIG) and prompt (TIG and PRS)
surface muon fluxes. Uncertainties in the conventional flux, unlike the prompt 
case, are 
not major, hence we have shown only the TIG parametrization. We note that 
depending on the flux model, the prompt fluxes rise above the conventional flux 
for (surface) muon energies between 200 TeV and 1000 TeV. In terms of (degraded)
muons entering the detector, we see from Figure \ref{fluxf} and Figure 
\ref{flux2} that 
this corresponds to measured muon energies of several tens of TeV and several 
hundreds of TeV. Thus, we note that underground muon measurements in this range 
will help reduce the present uncertainties in deducing the charm contributions to
muon and neutrino fluxes. Our calculations provide a
quantitative estimate of the feasibility and the  potential of these measurements to
accomplish this.
\begin{figure}[htp]
\hbox{\hspace{0cm}
\hbox{\includegraphics[scale=1]{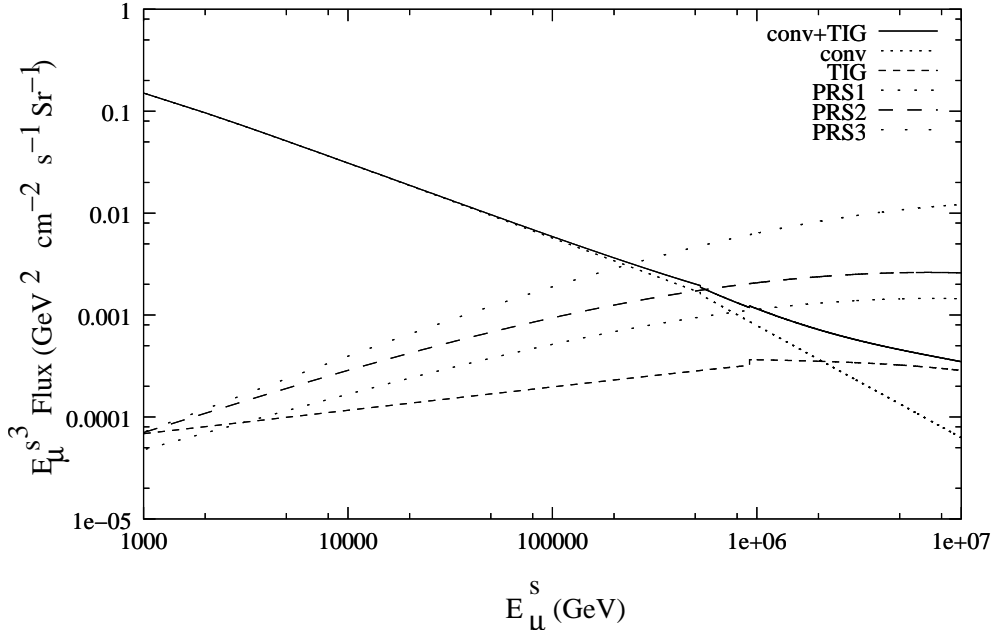}}}
\caption{(${E_{\mu}^s}^3 \times$  Flux) vs. surface muon enegy $E_\mu^s$ for
total TIG flux(solid), conventional(dotted), TIG prompt (short dashed), and 
prompt fluxes corresponding to PRS1(short spaced dots), PRS2(large dashed) and
PRS3(large spaced dots).}
\label{flux1}
\end{figure}
\begin{figure}[htp]
\hbox{\hspace{0cm}
\hbox{\includegraphics[scale=1]{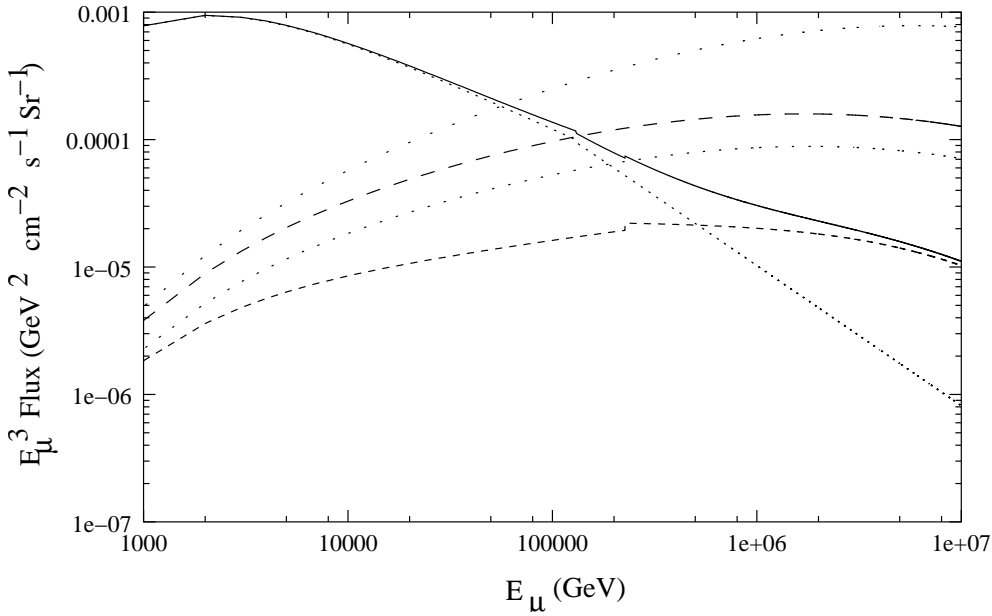}}}
\caption{($E_{\mu}^3 \times$  Flux) vs. energy of muon entering the
  underground detector $E_\mu$, ( for the  flux models listed
in the previous figure) after passing
through the rock distance of $3.5 \times 10^5 gm/cm^2.$}. Labelling of plots is identical to figure 
\ref{flux1}.
\label{flux2}
\end{figure}
\begin{figure}[htp]
\hbox{\hspace{0cm}
\hbox{\includegraphics[scale=1]{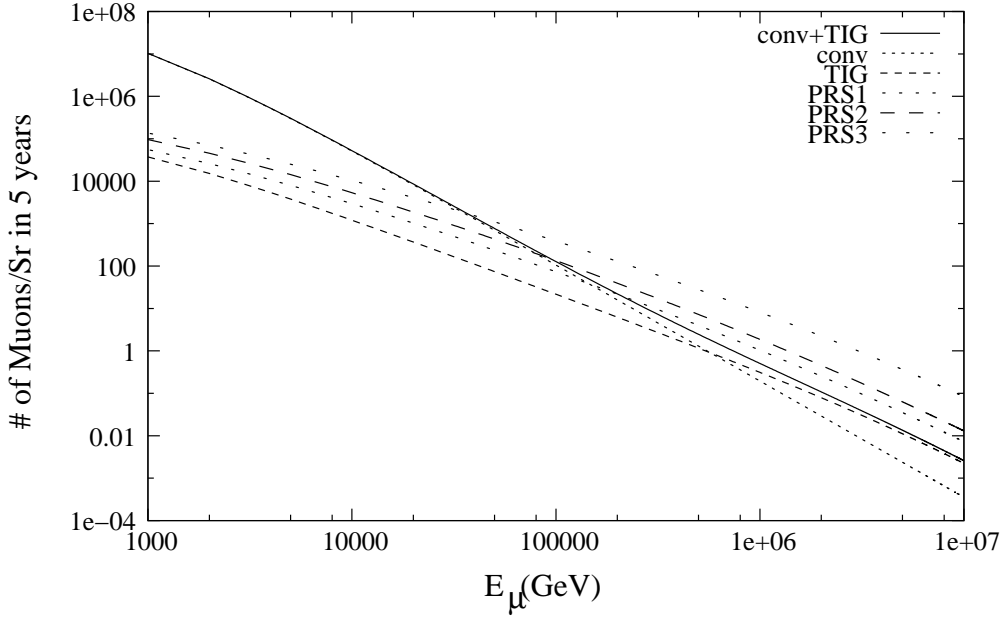}}}
\caption{Number of muons entering the 50 kT detector in 5 years per solid 
angle
vs. muon energy. Energy losses in the surrounding rock are taken into account. 
Depth of the detector is
assumed to be $3.5 \times 10^5 g/cm^2.$}
\label{nom}
\end{figure}

\section{Results and Discussion}
We are now in a position to calculate the expected cascade events for
a 50 kT detector in the energy range of interest discussed above.
While an entering muon in this energy range will produce observable
cascades, the number entering the detector over a given period is
limited by the sharply falling fluxes at these energies. It is thus
pertinent to  obtain a quantitative measure of this by estimating
$n_\mu$, the number of muons above a given threshold entering the 
detector per ster-radian for an exposure of $t$ years,
\begin{equation}
 n_{\mu} = \int_{E_{th}}^{\infty} dE_{\mu} \frac{dN}{dE_{\mu}} A 
\times t \ ,
\end{equation}
where A is the exposed area of a 50 kT iron detector.
This is shown in Figure \ref{nom} and Table \ref{tab3}. We note that while the 
number of entering muons for the lowest energy in Table \ref{tab3}, {\it i.e.} 
$1$ TeV is very large, one also obtains an observable number, {\it i.e.} 
1-3 events after integrating over solid angle  (considering that there
is no ``back-ground'' as such for such events) over the 5 year period
even for $E_\mu=1000$ TeV for the most conservative flux choice
(TIG). These energies delineate the muon energy range accesible. The number
of entering muons for all choices of PRS fluxes 
will be substantially higher, as shown. Even for the most conservative (TIG) 
flux choice, one expects good observational capability 
upto several hundred TeV\footnote{ We have given the per-steradian rates here. 
In order to predict the rate integrated over angle, 
predictions must carefully  take into account the depth dependance 
of the fluxes for a particular detector and its surrounding topography.}.

The number of cascades per muon  
above $E_0$=5, 10, 20, 50, 100, 300, 500, 1000, 5000 GeV respectively using Eq. \ref{ni} 
and 
\ref{cros} are tabulated in Table \ref{tab1}. 
Table \ref{tab1} also lists 
the surface muon energy $E_\mu^s$ corresponding to the underground muon energies 
$E_\mu$. While these are sample choices, it is clear that they can be 
further optimized based on the muon energy that one wants to 
observe to good statistical accuracy.  

The  {\it total} number of  cascade events per
ster-radian $N_c(E_0)$  (above  a given threshold $E_0$ and for a 50 kT $\times$ 5 yr exposure) is given by
\begin{equation}
N_c(E_0) = M(E_{\mu},E_0) n_{\mu}\ ,
\end{equation}
where $M(E_{\mu},E_0)$ is the cascade number calculated above in Section {\bf
2.1}. From Table \ref{tab3} and Table \ref{tab1}, we observe 
that this number is considerable for most thresholds, promising rich observational capabilities. 
For example, for the (most conservative) conventional+TIG flux model which we use as our benchmark, one finds
 that at even at  muon energies 
of 1000 TeV,one can produce 
51 events per solid angle for  a threshold of 5 GeV  and $\sim$1 event per solid angle for 
 a threshold of 5000 GeV. Expectedly, the  PRS models predict significantly more events compared to 
these estimates. 

 We have used the TIG flux as a benchmark to establish observability, since it 
leads to the most conservative event rate predictions. All other flux 
parametrisations lead to higher predictions. We note that even though TIG and PRS 
are not vastly different from each other in a qualitative sense since both are 
based on perturbative QCD inputs, their event rates in a large mass pair meter 
differ significantly. Indeed, the variations amongst fluxes in the same 
family(PRS1, PRS2, PRS3) are also large. 
Thus, the muon event rate can act as a soft  ( {\it i.e } not definitive, given the 
large uncertainties in the QCD predictions) discriminator between various prompt flux 
models and provide pointers to the physics input that should guide their 
development. Similarly, this rate provides a tool to better understand the 
present spectral uncertainities in the cosmic ray knee origin.  

\section{Conclusions}
Our main results are presented in Figs \ref{nom} and \ref{noi} and  
Tables \ref{tab3}, \ref{tab1}. From these we (conservatively) conclude  
that underground muon energy measurements for an energy 
range of $E_\mu$ of 1-1000 TeV are possible with a 50
kT iron detector \footnote{As stated earlier, we have used  a depth 
corresponding to the proposed INO site for specificity, however, the results can 
be easily generalised to other depths, as should be obvious.} running for 5
years. This will enable a better handle on the very high energy muon
fluxes between several TeV to about 5 PeV, and consequently illuminate our
estimates of the background muon and neutrino fluxes for ultra high
energy neutrino detectors and lessen present uncertainties in charm
production models. As emphasized earlier, the prompt muon flux is a measure 
of the prompt $\nu_e$ and $\nu_{\mu}$ flux, hence its importance to ultra 
high energy neutrino astronomy cannot be underestimated.

The observable muon energy range discussed in our results also corresponds 
to a range of 50 TeV to 50 PeV in {\it primary cosmic ray energies}. This 
range is crucial to an understanding of the origin of knee and our 
calculations demonstrate the feasibility and potential resulting from muon 
measurements for a better understanding of the origin of the knee. 

A detailed and comprehensive set of predictions for a given large mass 
detector necessarily requires a much more elaborate calculation of the muon 
losses than what is presented here, since local topography plays an 
important role in determining the surface muon energy corresponding to a 
measured muon energy. Our aim in this paper has been more to demonstrate 
observational feasibility rather than to make precise predictions.
Thus the calculations here show that 
very high energy muon measurements are possible in a large iron calorimeter 
and can aid in illuminating three important outstanding questions 
which address  partially overlapping isuues, one in cosmic 
ray physics, the second  in theoretical QCD and  and the 
third  in ultra high energy neutrino astronomy.

{\bf Acknowledgment:} We thank Namit Mahajan for helpful discussions. 
RG would also  like to thank Francis Halzen for informative  
discussions and John Beacom for pointing out some helpful references.

\end{document}